\documentclass[preprint,aps,12pt,showpacs,nofootinbib,tightenlines,amsmath,amssymb]{revtex4}
\usepackage{amsmath}
\usepackage{graphicx}
\usepackage{amssymb}
\usepackage{color}
\newcommand{\VEV}[1]{\left\langle #1\right\rangle}

\newcommand{\p}{\partial}

\newcommand{\MeV}{\;\text{MeV}}

\begin{document}
\def\intdk{\int\frac{d^4k}{(2\pi)^4}}
\def\sla{\hspace{-0.17cm}\slash}
\hfill

\title{ Infrared-Improved Soft-wall AdS/QCD Model for Mesons }

\author{Ling-Xiao Cui}\email{clxyx@itp.ac.cn}

\author{Zhen Fang}\email{fangzhen@itp.ac.cn}

\author{Yue-Liang Wu}\email{ylwu@itp.ac.cn}

\affiliation{State Key Laboratory of Theoretical Physics(SKLTP)\\
Kavli Institute for Theoretical Physics China (KITPC) \\
Institute of Theoretical Physics,
Chinese Academy of Sciences, Beijing 100190 \\
University of Chinese Academy of Sciences (UCAS), P.R. China }

\date{\today}

\begin{abstract}
We construct and investigate an infrared-improved soft-wall AdS/QCD model for mesons. Both linear confinement and chiral symmetry breaking of low energy QCD are well characterized in such an infrared-improved soft-wall AdS/QCD model. The model enables us to obtain a more consistent numerical prediction for the mass spectra of resonance scalar, pseudoscalar, vector and axial-vector mesons. In particular, the predicted mass for the lightest ground state scalar meson shows a good agreement with the experimental data. The model also provides a remarkable check for the Gell-Mann-Oakes-Renner relation and a sensible result for the space-like pion form factor.
\end{abstract}

\maketitle

\section{Introduction}
\label{Chap:Intro}

The property of asymptotic freedom of quantum chromodynamics (QCD) \cite{Gross:1973id,Politzer:1973fx} at short distances or high energies enables us to make perturbative treatment successfully on QCD at ultraviolet (UV) region.  At low energies, the QCD perturbative method is no longer applicable due to strong interactions at infrared(IR) region. So far we are still unable to solve from the first principle the low energy dynamics of QCD, while the chiral symmetry breaking and linear confinement are known to be two important features of non-perturbative QCD at the low energies. Many theoretical approaches have been developed to describe these two interesting properties of non-perturbative QCD. Such as the lattice QCD and effective quantum field theories. One often adopts effective quantum field theories to describe the low energy dynamics of QCD based on the approximate global chiral symmetry and dynamical chiral symmetry breaking \cite{Nambu:1960xd}. It has been shown explicitly in ref.\cite{DW} how to derive the spontaneous chiral symmetry breaking via the dynamically generated effective Higgs-like potential and obtain a consistent prediction for the mass spectra of the lowest lying nonet pseudoscalar and scalar mesons, where the scalars can be regarded as the composite Higgs-like bosons. While it is not manifest in such a chiral effective field theory to make predictions for the mass spectra of high resonance meson states.

The idea of large $N_c$ expansion \cite{Large Nc} and holographic QCD which has been explicitly realized via the anti-de Sitter/conformal field theory correspondence (AdS/CFT) \cite{Maldacena:1997re,Gubser:1998bc,Witten:1998qj} supply a new point of view for solving the challenging problem of strong interaction of QCD at low energies. The AdS/CFT establishes the duality between the weakly coupled supergravity in $AdS_5$ and the strongly coupled $\mathcal{N}=4$ super Yang-Mills gauge theory, which makes the calculations in strongly coupled theory feasible \cite{Polchinski:2001tt}. The so-called top-down and bottom-up approaches are two complementary methods in the way of pursuing the gravity/gauge duality of QCD. The former starts from some brane configurations in string theory to reproduce some basic features of QCD \cite{top down 1,top down 2,top down 3}. The latter which is known as AdS/QCD consists of bulk fields in a curved space to reproduce some experimental phenomena in QCD \cite{hard wall,DaRold:2005zs, soft wall, soft wall 1,unstable,Sui:2009xe,Li:2012ay}. It was also noticed in ref.\cite{BT1,BT2,BT3} that there exists the correspondence of matrix elements obtained in AdS/CFT with the corresponding formula by using the light-front representation. In this paper we will focus on the bottom-up approach.

There are the so-called hard-wall and soft-wall AdS/QCD models. The hard-wall AdS/QCD model was developed in \cite{hard wall,DaRold:2005zs}. The pattern of chiral symmetry breaking can be realized in the hard-wall AdS/QCD models, however the mass spectra for the excited mesons cannot match up with the experimental data well. In the simple soft-wall AdS/QCD model \cite{soft wall}, a dilaton background field with quadratic growth in the deep infrared(IR) region has been introduced to show the Regge behavior for the higher excited vector meson spectra. However, the chiral symmetry breaking phenomenon cannot consistently be realized in the simple soft-wall AdS/QCD model. This is because only introducing dilaton background field with quadratic growth at the IR region does not satisfy the equation of motion of the bulk field if other terms are not considered to be modified at the IR region. To incorporate the linear trajectories of QCD confinement and chiral symmetry breaking, nonlinear interactions of the bulk scalar have been considered in the previous studies \cite{Babington:2003vm,Casero:2007ae,Iatrakis:2010jb,Jarvinen:2011qe,unstable}. In \cite{Iatrakis:2010jb}, the meson spectrum was also studied phenomenologically, and a generalized model with back-reaction and finite number of flavors was considered in \cite{Jarvinen:2011qe}. In \cite{unstable}, a quartic interaction term was introduced to incorporate chiral symmetry breaking in the soft-wall model. However, the sign of quartic term leads to an unstable vacuum and results in a ghost ground state. To remedy this problem, in \cite{Sui:2009xe} the IR-deformed $AdS_5$ metric has been motivated to yield a consistent solution for the dilaton field from the equation of motion of the bulk field, which allows to introduce the quartic term with a correct sign for stabilizing the vacuum. The IR-deformed AdS/CFT metric structure is taken as the following simple form:
\begin{equation}\label{modified metric}
    ds^2=a^2(z)\left(\eta_{\mu\nu}dx^{\mu}dx^{\nu}-dz^2\right);\quad \mbox{with} \quad a^2(z)=(1+\mu_g^2z^2)/z^2
\end{equation}
where $\mu_g$ characterizes the confinement scale of low energy QCD. In the UV region $z\to 0$, the IR-deformed metric recovers conformal symmetry. It is interesting to observe that such a simply modified soft-wall AdS/QCD model \cite{Sui:2009xe}  can lead to a consistent solution for the dilaton background field from the equation of motion of the bulk field and meanwhile incorporate simultaneously both chiral symmetry breaking and linear confinement. In particular, the simple soft-wall AdS/QCD model with IR-deformed metric \cite{Sui:2009xe} can provide a consistent prediction for the mass spectra of resonance states for all the light scalar, pseudoscalar,vector and axial-vector mesons, except for the lightest ground state scalar meson which gets a much smaller mass in comparison with experimental data. In \cite{Sui:2010ay}, such a model has been extended to include three light flavor quarks.

As the IR-deformed metric does not satisfy Einstein equation, it causes inconvenience for considering the finite temperature effects \cite{finite T 1,finite T 2,finite T 3,finite T 4} as the ingoing wave boundary condition at the horizon of black hole cannot be satisfied directly. In the present paper, we are going to consider an alternative scheme to overcome such a shortage. Similar to considerations in ref.\cite{soft wall,Herzog:2006ra}, for simplicity, we do not consider the back-reaction of dilaton field to the gravity equation and take the dilaton field as a pure background field. There are a number of works\cite{Gubser:2008ny,Gursoy:2008za} in which the dilaton field and the gravitational field have been solved simultaneously from the background equations of motion so as to characterize the color effects of QCD. It is noticed that the IR-deformed metric may be converted into the original pure $AdS_5$ metric  by a scaling transformation of metric and bulk field redefinitions. As a consequence, the mass term of bulk scalar field has to be modified at the IR region, and the quartic interaction of bulk scalar field is necessary with a correct sign. Therefore, instead of taking IR-deformed metric, we are going to construct in the present paper an IR-improved AdS/QCD model for mesons. The z-dependence of the bulk mass term was suggested in \cite{z bulk mass1} from the running of operators, and it was studied in \cite{z bulk mass2,z bulk mass3} to be an alternative for incorporating chiral symmetry breaking. It was shown in \cite{z bulk mass2} that an unreasonable large quark mass is required to obtain the sensible mass spectra of mesons. In ref.\cite{z bulk mass3}, the scalar interaction was considered with a constant coupling. The mesons and nucleons mass spectra were calculated in \cite{peng zhang} with the consideration of both modified metric and bulk mass. However, in all considerations for the scalar meson part, the scalar states $f_0(980\pm10)$, $f_0(1505\pm6)$, $f_0(2103\pm8)$ and $f_0(2314\pm25)$ are incorrectly classified into the $SU(3)$ singlet resonance scalar states, they should belong to the isosinglet resonance scalar states of $SU(3)$ octet mesons. In this paper, we will introduce both the IR-modified bulk mass and bulk coupling of quartic scalar interaction to improve the situation and obtain a more consistent prediction for the mass spectra of resonance mesons.

The remaining parts of this paper are organized as follows. In Sec.\ref{Chap:Model}, a general IR-improved soft-wall AdS/QCD model is built. In Sec.\ref{Chap:Prediction}, we provide a numerical prediction for the mass spectra of the ground states and resonance states of all the light scalar, pseudoscalar, vector and axial-vector mesons with five appropriate model parameters. A discussion on the pion form factor is also presented. In Sec.\ref{Chap:Corrections}, we will present a detailed discussion on the possible influences and effects caused by the input parameters, such as  the quark mass, quark condensate, the scale $\mu_g$, parametrization of the coefficient of quartic interaction $\lambda_X$. The conclusions and remarks are presented in Sec.\ref{Chap:Sum}.

\section{IR-improved AdS/QCD Model for Mesons}
\label{Chap:Model}

In the AdS/QCD models, all fields are defined in a five-dimensional Anti-De Sitter space with the metric
\begin{equation}\label{metric}
    ds^2=a^2(z)\left(\eta_{\mu\nu}dx^{\mu}dx^{\nu}-dz^2\right);\qquad a^2(z)=\frac{1}{z^2} \, .
\end{equation}
The 5D action with quartic interaction term can be written as
\begin{equation}
S=\int d^{5}x\,\sqrt{g}e^{-\Phi(z)}\,{\rm {Tr}}\left[|DX|^{2}-m_{X}^{2} |X|^{2}-\lambda_X |X|^{4}-\frac{1}{4g_{5}^2}(F_{L}^2+F_{R}^2)\right] \, ,\label{action}
\end{equation}
with $D^MX=\p^MX-i A_L^MX+i X A_R^M$, $A_{L,R}^M=A_{L,R}^{M~a}t^a$ and ${\rm{Tr}}[t^at^b]=\delta^{ab}/2$. The gauge coupling $g_5$ is fixed to be $g_5^2 = 12\pi^2/N_c$ with $N_c$ the color number \cite{hard wall}. The complex bulk field $X$ will be decomposed into the  scalar and pseudoscalar mesons, the chiral gauge fields $A_L$ and $A_R$  will be identified to the vector and axial-vector mesons.

$\Phi(z)$ is the background dilaton field and its IR behavior must be quadratic to produce a linear trajectory of QCD confinement for excited meson masses \cite{soft wall}.  The introduction of background dilaton field destroys conformal symmetry and its asymptotic UV behavior must approach to zero to recover the conformal symmetry, i.e.,
\begin{equation}\label{dilatonBC}
    \Phi(z\rightarrow 0) \sim \mu _g^2 z^2 \sim 0; \qquad \Phi(z\rightarrow \infty)\sim \mu_g^2z^2 \  .
\end{equation}
For simplicity, we choose a simple parametrization for the dilaton field to satisfies the required IR and UV asymptotic behavior, namely,
\begin{equation}\label{dilaton}
    \Phi(z)=\mu_g^2z^2-\frac{\lambda_g^4\mu_g^4z^4}{(1+\mu_g^2z^2)^3}.
\end{equation}
with a constant parameter $\lambda_g$ which will be shown to cause the mass splitting between the ground state and the first excited meson state.

The quartic interaction of bulk scalar field also breaks the conformal symmetry, its coupling $\lambda_X$ is taken to be a z-dependent coupling, $\lambda_X =\lambda_X(z)$, and satisfy the following boundary conditions
\begin{equation}\label{lambdaBC}
   \lambda_X(z\rightarrow 0) \sim \mu _g^2 z^2 \simeq 0\ ; \qquad \lambda_X(z\rightarrow\infty)= \lambda \  .
\end{equation}
Here the vanishing UV behavior is chosen to keep the conformal symmetry of the Lagrangian at UV limit, while the choice of IR behavior can in general be arbitrary. In the previous papers\cite{unstable,Sui:2009xe,Sui:2010ay}  the coupling of quartic interaction term was taken to be a constant, here we only choose the IR behavior of $\lambda_X$ to be a constant for a simple consideration and also for a comparison with previous models. In fact, in the chiral effective field theory of low energy QCD in 4-dimensional space-time, the coupling of  quartic interaction term is a dimensionless constant. Thus we introduce the following parameterization for $\lambda_X(z)$
\begin{equation}\label{lambda}
    \lambda_X(z)=\frac{\mu _g^2 z^2}{1+\mu _g^2 z^2}\lambda\ .
\end{equation}
where $\lambda$ is a constant parameter. It will be shown in Sec.\ref{Chap:Corrections} that the resulting predictions for the mass spectra of resonance mesons  are  not sensitive to the specific forms of parameterization $\lambda_X(z)$.

Based on the AdS/CFT duality, the 5D conformal mass is given by $m^2_X = -3$ from the mass-dimension relation. The introduction of the background dilaton field and quartic interaction term causes the breakdown of the conformal symmetry in the IR region. As a consequence, when taking 5D AdS/CFT metric as a background, the mass term must get a corresponding modification in the IR region to yield a consistent solution for the equation of motion of the bulk scalar field $X$.  The IR-improved 5D conformal mass is expected to have the following boundary conditions for the leading terms
\begin{equation}\label{massBC}
   m^2_X(z\rightarrow 0) \sim -3 - O( \mu_g^2 z^2)  \simeq -3\ ; \qquad
   m^2_X(z\rightarrow \infty)\sim -\mu_m^2 z^2 - \lambda_m \  ,
\end{equation}
with the constant parameters $\mu_{m}$ and $\lambda_m$. The UV behavior of $m^2_X(z)$ is chosen to recover the 5D conformal mass. While the IR behavior is considered to deviate from the 5D conformal mass $m^2_X = -3$ with both the quadratic and constant corrections. These two terms are regarded as the leading order contributions indicated from the spontaneous chiral symmetry breaking via the dynamically generated effective composite Higgs-like potential\cite{DW}, where the mass term receives quadratic contributions with power-law running from the quark loops. With these considerations, we parameterize the IR-improved 5D conformal mass to have the following general form
\begin{equation}\label{mx}
    m_{X}^2(z)= -3-\frac{\lambda_1^2\mu_g^2z^2+\lambda_2^4\mu_g^4z^4}{1+\mu_g^2z^2} + \tilde{m}^2_X(z) \, ,
\end{equation}
with the constant parameters $\lambda_1$ and $\lambda_2$ as well as the z-dependent function $\tilde{m}^2_X(z)$. Here $\tilde{m}_X^2(z)$ represents the next-to-leading order contributions in both UV and IR sides. The above IR-improved 5D conformal mass will be determined by the equation of motion.

The expectation value of bulk scalar field $X$ is assumed to have a z-dependent form for two flavor case
\begin{equation}\label{VEV}
\VEV{X}=\frac{1}{2}v(z)\left(
                         \begin{array}{cc}
                           1 & 0 \\
                           0 & 1 \\
                         \end{array}
                       \right) \, .
\end{equation}
From the AdS/CFT duality, the bulk vacuum expectation value (bVEV) $v(z)$  has the following behavior at the UV boundary $z\to 0$:
 \begin{equation} \label{vzero}
 v(z\to0)=m_q \,\zeta\, z+\frac{\sigma\, z^3}{\zeta} \, ,
 \end{equation}
where $m_q$ and $\sigma$ are interpreted from the AdS/CFT duality and low energy QCD as the quark mass and quark condensate respectively. The normalization factor $\zeta$ is fixed by QCD with $\zeta=\sqrt{3}/(2\pi)$ ~\cite{z bulk mass1,zeta}.

To have the reasonable IR boundary condition for bVEV $v(z)$, it is interesting to notice the fact that the highly excited mesons exhibit parallel Regge trajectories as shown in \cite{Shifman:2007xn}, which indicates that the chiral symmetry is not restored with increasing excitation number. To incorporate the mass difference between vector and axial-vector resonances approaches to a constant as $z\to \infty$, the IR boundary condition for the bVEV $v(z)$ is expected to be linear
\begin{equation} \label{vinfty}
 v(z\to \infty)= v_q  z \, ,
 \end{equation}
with $v_q$ the constant parameter which characterizes the energy scale of dynamically generated spontaneous chiral symmetry breaking caused by the quark condensate\cite{DW}. To realize the above UV and IR behaviors, we simply parameterize the bVEV  of bulk scalar field as follows
\begin{equation}\label{bVEV}
    v(z)=\frac{A z+B z^3}{1+C z^2} \, ,
\end{equation}
with
\begin{equation}\label{ABC}
     A=m_q\zeta,\quad B=\frac{\sigma}{\zeta}+m_q\zeta C,\quad C=\mu_c^2/\zeta,\quad v_q =B/C \, ,
\end{equation}
where the constant parameter $\mu_c$ characterizes the QCD confinement scale.

Actually, the IR-behavior of bVEV $v(z)$ in the boundary $z\to\infty$ is correlated to the behavior of the background dilaton field given in Eq.~(\ref{dilatonBC})  and the quartic interaction term given in Eq.~(\ref{lambdaBC}) as well as the IR-improved 5D conformal mass given in Eq.~(\ref{massBC}). To see that, let us analyze from the bulk action the equation of motion of the bVEV $v(z)$, which correlates the background dilaton $\Phi(z)$, the IR-improved 5D conformal mass $m_X^2(z)$ and the quartic interaction term $\lambda_X(z)$ of bulk scalar field
\begin{equation}\label{equa of VEV}
\partial_z \left(a(z)^3 e^{-\Phi (z)}\partial_{z}v(z)\right)-a(z)^5
   e^{-\Phi (z)}
   \left(m_X^2(z)+\frac{1}{2} \lambda_X (z)
   v(z)^2\right) v(z)=0 \, .
\end{equation}
Its general solution for the IR-improved 5D conformal mass is given by
\begin{equation}\label{solve mx}
   m_X^2(z)=-\frac{1}{2} \lambda_X(z)  v(z)^2-\frac{\Phi '(z)
   v'(z)}{a(z)^2 v(z)}+\frac{3 a'(z)
   v'(z)}{a(z)^3 v(z)}+\frac{v''(z)}{a(z)^2
   v(z)} \, .
\end{equation}
From the parametrization given in Eqs.~(\ref{dilaton},\ref{lambda},\ref{bVEV}),  the UV behavior of IR-improved 5D conformal mass is obtained to be
\begin{equation}\label{mx UV}
    m_X^2(z\rightarrow0)\to -3-2\mu_g^2z^2 \simeq -3 \, ,
\end{equation}
which leads to the UV limit $m^2_X=-3$ given by the mass-dimension relation $\Delta(\Delta-4)=m_X^2$. The IR behavior is found to have the following general form
\begin{equation}\label{mx IR}
    m_X^2(z\rightarrow\infty)\simeq -\left(\frac{B^2\lambda}{2C^2}+2\mu_g^2\right)z^2+ \left(-3+\frac{B^2\lambda}{2C^2\mu_g^2} +\frac{(B-AC)}{C} ( \frac{B\lambda}{C^2}- \frac{4\mu_g^2}{B} )\right) \, .
\end{equation}
After matching the leading order terms to the IR-improved 5D conformal mass formalism given in Eq.~(\ref{mx}),
the two constant parameters $\lambda_1$ and $\lambda_2$  are completely determined to be
\begin{eqnarray}
  \lambda_1 &=& \lambda_2 = \sqrt{2} \, , \quad \mbox{or} \quad \mu_m = 2\mu_g\, , \quad \lambda_m = 1
\end{eqnarray}
and the VEV $v_q$ due to quark condensate is given via the following relation
\begin{eqnarray}
 v_q=  \frac{B}{C} =  \frac{\sigma }{\mu_c^2} + m_q  \zeta = \sqrt{\frac{(2\mu_g)^2}{\lambda} }= \sqrt{\frac{\mu_m^2}{\lambda} } \, ,
\end{eqnarray}
which shows the familiar formalism of VEV in the Higgs-like mechanism. Such a result is consistent with the minimal condition of dynamically generated composite Higgs-like potential in the chiral dynamical model when the chiral symmetry is broken down spontaneously\cite{DW} . The IR behavior of 5D conformal mass is found to be
\begin{equation}\label{massIR}
   m^2_X(z\rightarrow \infty)\simeq -(2\mu_g)^2 z^2 -1 \  .
\end{equation}

So far we have described the IR-improved soft-wall AdS/QCD for mesons with appropriate modifications in the IR region for the background dilaton field and the 5D conformal mass and quartic interaction coupling of bulk scalar fields.

\section{Numerical Predictions}
\label{Chap:Prediction}

 With the above analyses on the IR-improved AdS/QCD model for mesons, we are now in the position to make a numerical predication for the mass spectra of resonance mesons.  For that, we shall first determine the five free parameters involved in the model, they are $m_q$, $\sigma$, $\mu_c$, $\mu_g$ and $\lambda_g$.

\subsection{Input Parameters}

The three parameters $m_q$, $\sigma$ and $\lambda_g$ are directly fixed by the well-known experimental values of  the $\pi$ meson mass $m_{\pi}=139.6\MeV$ and the $\pi$ meson decay constant $f_{\pi}=92.4\MeV$ as well as the $\rho$ meson mass $m_{\rho}=775.5\MeV$. The energy scale parameters $\mu_g$ and $\mu_c$ are determined by taking a global fitting of the slopes for the resonance vector and pseudoscalar mesons.  The pion decay constant is given by the axial-vector equation of motion with the pole in the propagator set to zero as showed in \cite{hard wall}.
\begin{equation}\label{fpi}
    f_{\pi}^2=\left.-\frac{1}{g_5^2}\frac{\partial_zA(0,z)}{z}\right|_{z\rightarrow0}
\end{equation}
$A(0,z)$ is the axial-vector bulk-to-boundary propagator which satisfies the boundary conditions $A(0,0)=1$ and $\partial_zA(0,z\rightarrow\infty)=0$. The values of four fitting parameters are presented in the Table \ref{Table:parameter}, it shows that the three scales are at the range $300\sim 500$ MeV which is around the QCD scale $\Lambda_{QCD}$.

\begin{table}[!h]
\begin{center}
\begin{tabular}{ccccc}
\hline\hline
 $\lambda_g$ &$m_q$(MeV) & $\sigma^{\frac{1}{3}}$(MeV) & $\mu_g$(MeV)  & $\mu_c$(MeV) \\
\hline
1.7 & 3.52 & 290 & 473 & 375   \\
\hline\hline
\end{tabular}
\caption{The values of five parameters}
\label{Table:parameter}
\end{center}
\end{table}

\subsection{Pseudoscalar Mesons}

The bulk scalar field can be written in the form of fluctuation fields $X(x,z)\equiv(v(z)/2+S(x,z))e^{2i\pi(x,z)}$, with $S(x,z)$ being the scalar meson field and $\pi(x,z)=\pi^a(x,z) t^a$ being the pseudoscalar meson field. Kaluza-Klein decomposition is preferred to find mass eigenvalues $\pi(x,z)=\sum_n\varphi_n(x)\pi_n(z)$. In order to cancel the cross term of axial-vector and pseudo-scalar fields, we decompose the axial field in terms
of its transverse and longitudinal components, $A_{\mu}^{a}=A_{\mu \bot}^{a}+\partial_{\mu}\phi^{a} $. By choosing the axial gauge $A_z=0$, we then obtain the following equation of motion in the 4D momentum space from the action\,(\ref{action}).
\begin{eqnarray}
\label{equa of pi}
& &\partial_z\left(a(z)e^{-\Phi}\partial_z\phi^{a}\right)+g_5^2\,a^3(z)\,v^2(z)e^{-\Phi}\,(\pi^{a}-\phi^{a})=0\\
& &q^2\partial_z\phi^{a}-g_5^2\,a^2(z)\,v^2(z)\partial_z\pi^{a}=0
\end{eqnarray}
The bound state modes in the bulk correspond to the hadrons of QCD. The eigenvalue of pseudoscalar meson mass can be found by using the boundary conditions $\pi(z\rightarrow0)=0$, $\partial_z\pi(z\rightarrow\infty)=0$. After eliminating the longitudinal component, and by adopting the shooting method, we arrive at  the numerical predictions for the resonance pseudoscalar mesons given in Table \ref{Table:PS mass}. The results show a good agreement between theoretical predictions and experimental data.
\begin{table}
\begin{center}
    \begin{tabular}{|c|c|c|c|c|c|c|c|}
       \hline
       $\pi$ & 0 & 1 & 2 & 3 & 4 & 5 & 6 \\
       \hline\hline
       Exp.(MeV) & 139.6 & $1300\pm100$ & $1812\pm12$ & $2070\pm35^\dag$ & $2360\pm25^\dag$ & --- & --- \\
       \hline
       Theory (MeV) & $139.6^* $& 1465 & 1813 & 2068 & 2287 & 2482 & 2662 \\
       \hline
     \end{tabular}
\end{center}
\caption{The experimental and predicted mass spectra for pseudo-scalar mesons. $\dag$ -appears strictly in the further states of \cite{PDG}. }
\label{Table:PS mass}
\end{table}

\subsection{Scalar Mesons}

By expanding the 5D scalar field as $S(x,z)=\sum_n\psi_n(x)S_n(z)$, we obtain from the equation of motion of the bulk scalar the following eigenvalue equation for the resonance scalar mesons:
\begin{equation}\label{equa of S}
     \p_z\left(a^3(z)e^{-\Phi}\p_z S_n(z)
  \right)-a^5(z)e^{-\Phi}m_X^2S_n(z)=-a^3(z)e^{-\Phi}m_{S_n}^2S_n(z)
\end{equation}
Here we only consider the $SU(3)$ singlet scalar mesons. It has been discussed in \cite{Sui:2009xe} that the scalar states $f_0(980\pm10)$,$f_0(1505\pm6)$,$f_0(2103\pm8)$ and $f_0(2314\pm25)$ should be classified into the isosinglet resonance scalar states of $SU(3)$ octet mesons, rather than the $SU(3)$ singlet resonance states.
Under the boundary conditions $S_n(z\rightarrow0)=0$, and $\partial_zS_n(z\rightarrow\infty)=0$, by applying the shooting method to solve Eq. (\ref{equa of S}), we can obtain the numerical predictions for the mass spectra of $SU(3)$ singlet resonance scalar mesons, the results are presented in Table \ref{Table:S mass}. It is seen that the theoretical predictions agree remarkably with the experimental data. Note that our input parameters do not use any data involving scalar mesons.
\begin{table}
\begin{center}
   \begin{tabular}{|c|c|c|c|c|c|c|c|}
     \hline
     $f_0$ & 0 & 1 & 2 & 3 & 4 & 5 & 6 \\
     \hline\hline
     Exp.(MeV) & $400-550$ & $1200-1500$ & $1720 \pm 6$ & $1992 \pm 16$ & $2189 \pm 13$ & --- & --- \\
     \hline
     Theory (MeV) & 460 & 1359 & 1723 & 1989 & 2216 & 2417 & 2602 \\
     \hline
   \end{tabular}
\end{center}
\caption{The experimental and predicted mass spectra for scalar mesons}
\label{Table:S mass}
\end{table}

\subsection{Vector Mesons}

The equation of motion for the vector part can be derived from the action (\ref{action}) by choosing the axial gauge $V_5=0$
\begin{equation}\label{equa of V}
    \partial_z(e^{-\Phi}a(z)\partial_zV_n(z))+m_{V_n}^2e^{-\Phi}a(z)V_n(z)=0\ ,
\end{equation}
which can also be solved by using the shooting method. With the boundary conditions $V_n(z\rightarrow0)=0$, and $\partial_zV_n(z\rightarrow\infty)=0$, we easily obtain the solutions for the above vector meson eigenvalue equation and arrive at the numerical predictions for the mass spectra of resonance vector mesons. The results are presented in Table \ref{Table:V mass}, which shows a good agreement in comparison with the experimental data.
\begin{table}
\begin{center}
    \begin{tabular}{|c|c|c|c|c|c|c|c|}
       \hline
       $\rho$ & 0 & 1 & 2 & 3 & 4 & 5 & 6 \\
       \hline\hline
       Exp.(MeV) & $775.5\pm0.34$ & $1460\pm25$ & $1720\pm20$ & $1909\pm30$ & $2149\pm17$ & $2265\pm40$ & --- \\
       \hline
       Theory (MeV) & 775.5* & 1354 & 1667 & 1919 & 2140 & 2339 & 2523 \\
       \hline
     \end{tabular}
\caption{The experimental and predicted mass spectra for vector mesons}
\label{Table:V mass}
\end{center}
\end{table}

\subsection{Axial-vector Mesons}

Minimizing the action (\ref{action}) with the axial gauge $A_5=0$, we derive the axial meson field equation of motion for its perpendicular component $A_{\mu\bot}^{a}$ which is abbreviated as $A_n$ below.
\begin{equation}\label{equa of AV}
    e^{\Phi}\partial_z(a(z)e^{-\Phi}\partial_zA_{n})+a(z)m_{A_a}^2A_{n}-a^3(z)g_5^2v^2(z)A_{n}=0
\end{equation}
Again, with the boundary conditions $A_n(z\rightarrow0)=0$, and $\partial_zA_n(z\rightarrow\infty)=0$, we can solve the above eigenvalue equation by using the shooting method and yield the numerical predictions for the mass spectra of resonance axial-vector mesons. The results are shown in Table \ref{Table:AV mass}, a good global fitting with experimental data is also obtained, except for the ground state meson mass which is slightly smaller than the experimental result.
\begin{table}
\begin{center}
    \begin{tabular}{|c|c|c|c|c|c|c|c|}
       \hline
       $a_1$ & 0 & 1 & 2 & 3 & 4 & 5 & 6 \\
       \hline\hline
       Exp.(MeV) & $1230\pm40$ & $1647\pm22$ & $1930^{+30}_{-70}$ & $2096\pm122$ & $2270^{+55}_{-40}$ & --- & --- \\
      \hline
       Theory(MeV) & 1095 & 1626 & 1913 & 2145 & 2350 & 2537 & 2709 \\
       \hline
     \end{tabular}
\caption{The experimental and predicted mass spectra for axial-vector mesons}
\label{Table:AV mass}
\end{center}
\end{table}

\subsection{Pion Form Factor}

It has been shown that the above IR-improved soft-wall AdS/QCD model with IR-modified 5D conformal mass and quartic interaction can provide a remarkable prediction for the mass spectra of all light resonance mesons. Here we shall calculate the pion form factor $F_\pi(q^2)$ for a further consistent check on such an IR-improved soft-wall AdS/QCD model. The space-like pion form factor can be determined from the sum over vector meson poles
\begin{equation}\label{form factor 1}
    F_{\pi}(q^2)=-\sum_{n=1}^{\infty}\frac{f_ng_{\rho_n\pi\pi}}{q^2-m_{V_n}^2}
\end{equation}
where $f_n$ is the decay constants of the vector mesons. For a practical calculation, we adopt the expression in terms of the vector and axial-vector bulk-to-boundary propagators \cite{form factor},
\begin{equation}\label{form factor}
     F_\pi(q^2)=\frac{g_{5}}{N}\int dz\,
 V(q,z)\,e^{-\Phi(z)}\left(\frac{a(z)(\partial_z\varphi(z))^2}{g_{5}^2}+v^2(z)a^3(z)(\pi(z)-\varphi(z))^2\right)
\end{equation}
where the $V(q,z)$ is the vector bulk-to-boundary propagator with the boundary conditions $V(q,z\rightarrow0)=1$ and $V(0,z)=1$. The functions $\pi(z)$ and $\varphi(z)$ are the solutions of Eq.(\ref{equa of pi}) with the normalization
\begin{equation}\label{normalized pi}
 N= \int dz\, e^{-\Phi(z)}\,\left(\frac{a(z)(\partial_z\varphi(z))^2}{g_{5}^2}+v^2(z)a^3(z)(\pi(z)-\varphi(z))^2\right)
\end{equation}
The integration region is the whole range of $z\in(0,\infty)$. While in the numerical calculation the upper limit is chosen to obtain a near zero integrand due to the suppressed factor $e^{-\Phi(z)}$, the lower limit should be assigned a value as small as possible to obtain a stable result.

\begin{figure}[ht]
\begin{center}
\includegraphics[width=10cm,clip=true,keepaspectratio=true]{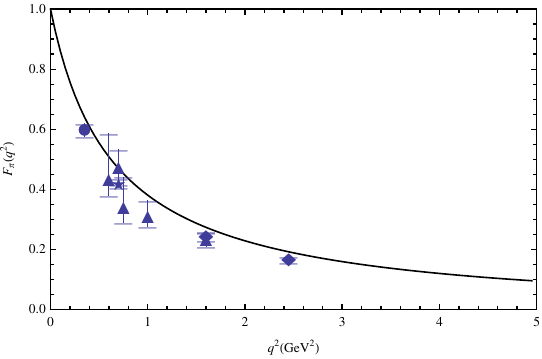}
\caption{The solid line shows the space-like behavior of pion form factor $F_\pi(q^2)$ predicted in the IR-improved AdS/QCD model for mesons, which is compared with the experimental data analyzed in \cite{form factor}. The triangles are data from DESY, reanalyzed in \cite{form factor 1}. The diamonds are data from Jefferson Lab \cite{form factor 2}. The circles \cite{form factor 3} and the stars \cite{form factor 4} are data obtained from DESY}
\label{fig:form factor}
\end{center}
\end{figure}

The numerical result is plotted in Fig. \ref{fig:form factor}, which shows that the IR-improved AdS/QCD model for mesons can provide a quite well description on the pion form factor.

\section{Influences of Input Parameters}
\label{Chap:Corrections}

  From the above analyses and numerical results, we see a good agreement between theoretical predictions and experimental data with five input parameters. It is then intriguing to ask how the input parameters influence the predictions.

\subsection{Effects of different IR behavior and parameterization of $\lambda_X(z)$}
\label{Chap:lambda}

We now discuss the possible effects caused from different parameterizations of the coupling of quartic interaction $\lambda_X(z)$. In general, there are various parameterizations for the z-dependent function to fit the required UV and IR boundary conditions.  For the coupling of the quartic interaction $\lambda_X(z)$, we have made a simple parametrization given in Eq. (\ref{lambda}). Here we shall consider other typical parameterizations with the same IR behavior to see the possible influences. Three other parametrization forms for $\lambda_X(z)$ and their IR behavior are showed in Table.\ref{Table:forms of lambda}.
\begin{table}
\begin{center}
    \begin{tabular}{|c||c|c|}
       \hline
       Three Cases &  Parametrization Forms of $\lambda_X(z)$ &IR behavior of $\lambda_X(z)$\\
       \hline\hline
      A &  $    \lambda_X(z)=(\sqrt{1+\lambda_g^2z^2}-1)/\sqrt{1+\mu_g^2z^2}$ & $\lambda_X(z\to\infty)\sim constant$ \\
      \hline
      B &  $\lambda_X(z)=\lambda_g^4 z^4/(1+\mu_g^2 z^2)^2$ &  $\lambda_X(z\to\infty)\sim constant$\\
      \hline
      C & $\lambda_X(z)=constant$ &  $\lambda_X(z\to\infty)\sim constant$ \\
      \hline
     \end{tabular}
\caption{Different parameterizations of $\lambda_X(z)$ and their IR behaviors. In cases A and B, the UV behaviors are the same $\lambda_X(z\to 0)\sim 0$. In case C, a constant coupling $\lambda_X$ is taken for a comparison.}
\label{Table:forms of lambda}
\end{center}
\end{table}

In cases A and B, the UV behaviors are the same with $\lambda_X(z\to 0)\sim 0$, so as to recover the conformal symmetry at UV limit. The IR behavior of cases A and B is consistent with Eq. (\ref{lambdaBC}). In case C we take a constant coupling $\lambda_X$ for a comparison as it has been considered in all previous works. The corresponding numerical results for these three cases are presented in Table.\ref{Table:lambda results}. Note that the quartic interaction of bulk scalar field can only affect the mass spectra of scalar mesons.

\begin{table}
\begin{center}
   \begin{tabular}{|c|c|c|c|c|c|c|c|}
     \hline
     $f_0$ & 0 & 1 & 2 & 3 & 4 & 5 & 6 \\
     \hline\hline
     Exp.(MeV) & $400-550$ & $1200-1500$ & $1720 \pm 6$ & $1992 \pm 16$ & $2189 \pm 13$ & --- & --- \\
     \hline
     Case A (MeV) & 495 & 1363 & 1720 & 1984 & 2209 & 2410 & 2594 \\
     \hline
     Case B (MeV) & 446 & 1397 & 1786 & 2060 & 2289 & 2490 & 2674 \\
     \hline
     Case C (MeV) & 443 & 1305 & 1655 & 1919 & 2146 & 2348 & 2534 \\
     \hline
   \end{tabular}
\end{center}
\caption{Mass spectra of scalar mesons with different parametrization for $\lambda_X(z)$}
\label{Table:lambda results}
\end{table}

From Table.\ref{Table:lambda results}, one can see that the resulting predictions for the mass spectra of resonance scalar mesons are not sensitive to the specific forms of parametrization. In all cases A, B and C, the appropriate IR behavior leads to a consistent prediction for the scalar mesons. Note that the sign of the quartic interaction is crucial to obtain a stable vacuum. An opposite sign of quartic term was taken in ref. \cite{unstable}, which yielded an unstable vacuum and resulted in a ghost ground state.

\subsection{Influences of Parameters $m_q$, $\mu_g$ and $\lambda_g$}
\label{Chap:error}

\begin{figure}[ht]
\begin{center}
\includegraphics[width=10cm,clip=true,keepaspectratio=true]{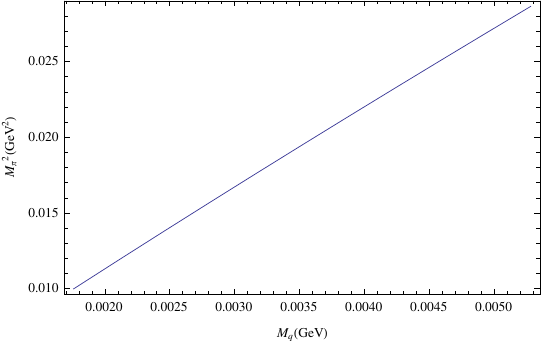}
\caption{The $m_q$-$m_\pi^2$ relation as a consistent check of the GOR relation}\label{fig:GOR}
\end{center}
\end{figure}

Let us now consider how the variation of parameters ($m_q$, $\mu_g$ and $\lambda_g$) influences the theoretical predictions for the mass spectra of resonance mesons. The extracted quark mass  $m_q =3.52$ MeV from the IR-improved AdS/QCD model for mesons agrees remarkably  with the one $m_q$ = $\overline{m}_{ud}=(3.50^{+0.7}_{-0.2})\MeV$ cited in \cite{PDG} as the mean value of u-quark and d-quark for the two flavour case. It can be shown that the quark mass has a very little influence on the mass spectra except for $\pi$ mass. We plot the $m_q$-$m_\pi^2$ relation below Fig.\ref{fig:GOR}, which shows a perfect linear relation and provides a consistent check on the Gell-Mann-Oakes-Renner (GOR) relation $f_\pi^2m_\pi^2=2m_q\sigma$. Actually, it is interesting to note that one can derive the GOR relation from the present model just following the procedure of \cite{hard wall}.

\begin{figure}[!h]
\begin{center}
\includegraphics[width=64mm,clip]{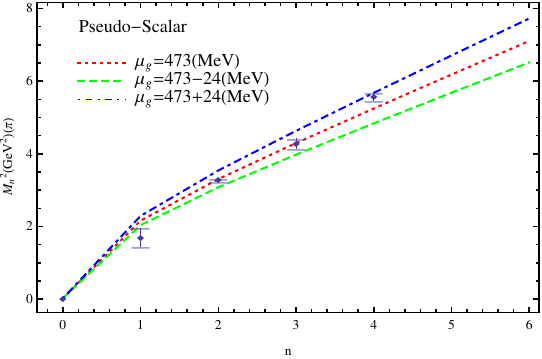}
\includegraphics[width=64mm,clip]{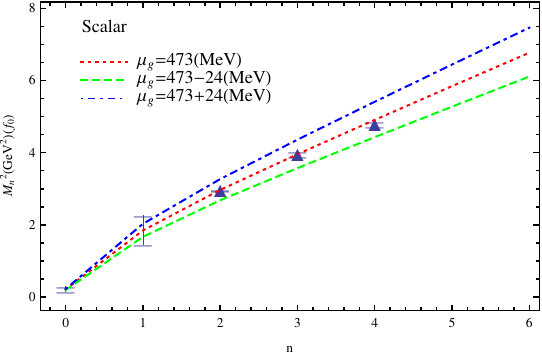}
\includegraphics[width=64mm,clip]{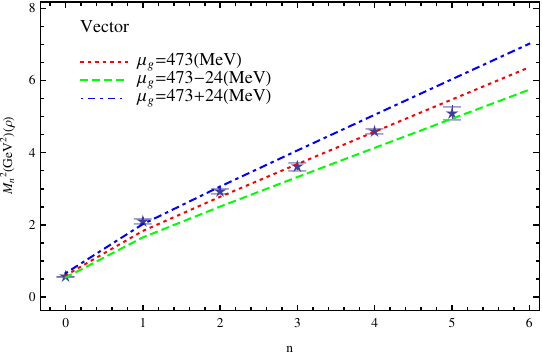}
\includegraphics[width=64mm,clip]{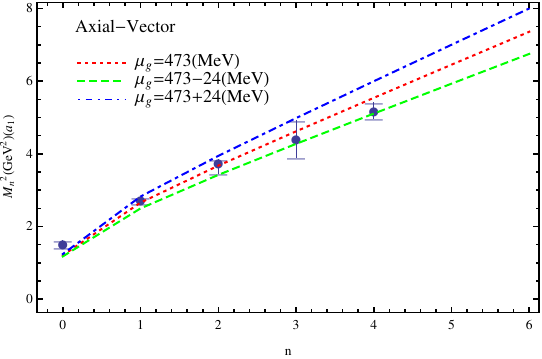}
\end{center}
\caption{The influences of scale $\mu_g$ in pseudo-scalar mesons(top,left), scalar mesons(top right), vector mesons(bottom,left) and axial-vector(bottom, right)}\label{Fig mug}
\end{figure}

\begin{figure}
\begin{center}
\includegraphics[width=64mm,clip]{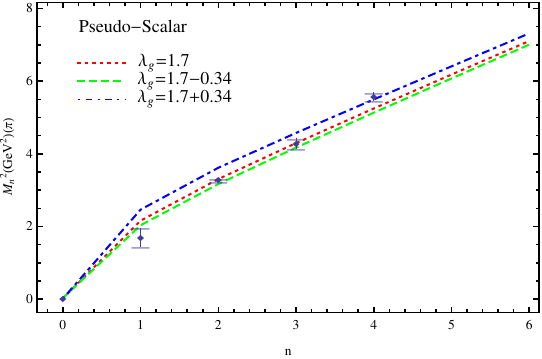}
\includegraphics[width=64mm,clip]{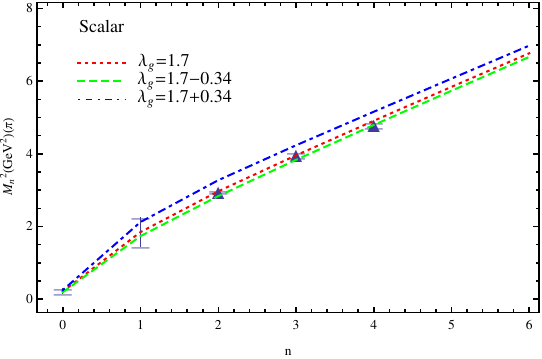}
\includegraphics[width=64mm,clip]{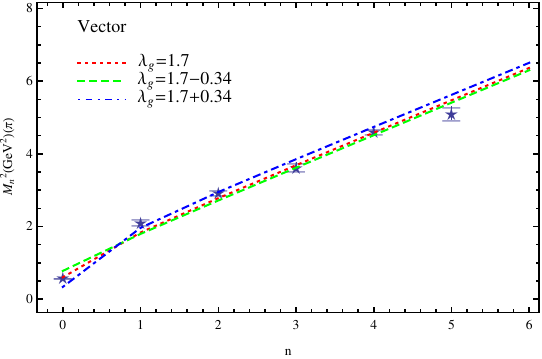}
\includegraphics[width=64mm,clip]{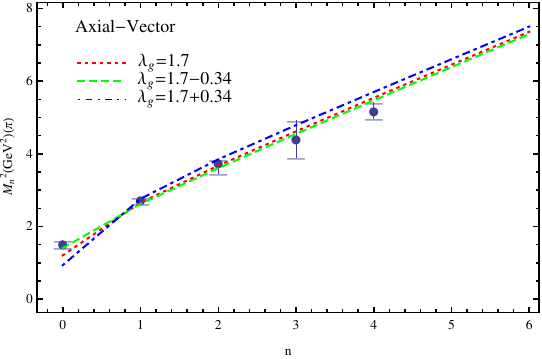}
\end{center}
\caption{The influences of parameter $\lambda_g$ in pseudo-scalar mesons(top,left), scalar mesons(top right), vector mesons(bottom,left) and axial-vector(bottom, right)}\label{Fig lambda}
\end{figure}

The influences arising from the model parameters $\mu_g$ and $\lambda_g$ have been studied and plotted in Fig.\ref{Fig mug} and Fig.\ref{Fig lambda}. One can see that the scale parameter $\mu_g$ mainly governs the average slope of mass spectra, while the parameter $\lambda_g$ appearing in the next-to-leading term of background dilaton field plays a very important role in manifesting the large gap between the ground state and the first excited state. Meanwhile, the parameter $\lambda_g$ also brings a significant effect in a global lifting of the highly excited meson masses.

\newpage

\section{Conclusions and Remarks}
\label{Chap:Sum}

       The soft-wall AdS/QCD with IR-deformed 5D metric was shown to provide a consistent prediction for the mass spectra of all light resonance mesons \cite{Sui:2009xe} except for the lightest ground state scalar meson, while it is very crucial to understand the property of the lightest ground state scalar meson as it codes the information of dynamically spontaneous chiral symmetry breaking mechanism\cite{DW}.  This comes to our main motivation to explore an alternative IR-improved soft-wall AdS/QCD model in the present paper. Also as the IR-deformed 5D metric causes an inconvenience for studying the finite temperature effects due to a complicated dilaton solution\cite{finite T 1,finite T 2,finite T 3,finite T 4}, it motivates us to make a scaling transformation to convert the IR-deformed 5D metric back to the original 5D AdS/CFT metric, and attribute its effects into the IR-improved soft-wall AdS/QCD model  with the IR-modified 5D conformal mass and IR-modified quartic interaction.

       In the IR-improved soft-wall AdS/QCD model constructed in this paper, the UV asymptotic behavior of all the IR-modified quantities has been set to maintain the conformal symmetry, which is the required feature of QCD at short distance or high energy. The background dilaton field has been parameterized to characterize the linear confinement of QCD by a simple IR-modified form with two parameters,  both IR and UV asymptotic behaviors have been chosen to be quadratic and keep the same asymptotic form as suggested in \cite{Sui:2009xe} for the better fitting reason. The bVEV of bulk scalar field has the well-known UV behavior due to AdS/CFT duality, while the IR behavior of the bVEV is not manifest, we have taken a linear form as indicated from the dynamically generated spontaneous chiral symmetry breaking  due to quark condensate\cite{DW}. Since the quartic interaction of bulk scalar field spoils the conformal symmetry, so that the UV behavior of its coupling has been set to vanish $\lambda_X(z\to 0) = 0$, and the IR behavior is assumed to approach a constant $\lambda_X(z\to \infty) = \lambda$. The UV asymptotic behavior of the IR-modified 5D conformal mass is given to be $m_X^2(z\to 0) =-3$ via the mass-dimension relation due to AdS/CFT duality, the leading IR behavior of 5D conformal mass has been taken to be similar to the one of background dilaton and grow in a quadratic form as indicated from the dynamically generated composite Higgs-like potential in the chiral dynamical model\cite{DW}.  It has been shown that the IR-modified quantities in the action are actually correlated via the equation of motion of bulk scalar field, the leading IR and UV asymptotic behaviors of all the IR-modified quantities must be matched consistently.  In the practical construction of the model, we have parameterized the IR-modified forms for the background dilaton, the coupling of quartic interaction, the bVEV of bulk scalar field and the leading 5D conformal mass. The general solution for the IR-modified 5D conformal mass is considered to be determined by solving the equation of motion of bulk scalar field.

   We have demonstrated that such an IR-improved soft-wall AdS/QCD model for mesons can well describe the main features of QCD with linear confinement and  chiral symmetry breaking. The model containing five parameters has been shown to provide a consistent prediction for the mass spectra of resonance scalar, pseudoscalar, vector and axial-vector mesons. The agreement between the theoretical predictions and experimental data can remarkably be obtained, and the deviations have been shown to be with a few percent. In particular, the ground state mass of the lightest scalar meson has been raised to reach a good agreement with the experimental data.  The model has led to a perfect GOR relation and a sensible space-like pion form factor $F_\pi(q^2)$ . We have also analyzed the effects of input parameters on the numerical predictions for the mass spectra of mesons, and shown how the input parameters influence the slope of linear trajectories of mass spectra and the large gap between the ground state and the first excited resonance state. It has been demonstrated that the five parameters in the model can well be extracted from the five precisely measured experimental inputs. Finally, we would like to point out that the gravity is treated as the background, the back reacted effect is not included as it is expected to be insignificant\cite{BR}.

\section*{Acknowledgements}
The authors would like to thank Y.B. Yang for useful discussions. This work is supported in part by the National Nature Science Foundation of China (NSFC) under Grants No. 10975170, No.10905084,No.10821504; and the Project of Knowledge Innovation Program (PKIP) of the Chinese Academy of Science.


\end{document}